\documentclass[11pt]{revtex4}                   %revised 9 May 2012%
\usepackage{amssymb,epsf}
\usepackage{latexsym}

\begin{document}

\title{Higher dimensional charged $f(R)$ black holes}
\author{Ahmad Sheykhi \footnote{sheykhi@uk.ac.ir}}
\address{Center for Excellence in Astronomy and Astrophysics (CEAA-RIAAM) Maragha, P. O. Box 55134-441, Iran\\
         Physics Department and Biruni Observatory, Shiraz University, Shiraz 71454, Iran }

\begin{abstract}
We construct a new class of higher dimensional black hole
solutions of $f(R)$ theory coupled to a nonlinear Maxwell field.
In deriving these solutions the traceless property of the
energy-momentum tensor of the matter filed plays a crucial role.
In $n$-dimensional spacetime the energy-momentum tensor of
conformally invariant Maxwell field is traceless provided we take
$n=4p$, where $p$ is the power of conformally invariant Maxwell
lagrangian. These black hole solutions are similar to higher
dimensional Reissner-Nordstrom AdS black holes but only exist for
dimensions which are multiples of four. We calculate the conserved
and thermodynamic quantities of these black holes and check the
validity of the first law of black hole thermodynamics by
computing a Smarr-type formula for the total mass of the
solutions. Finally, we study the local stability of the solutions
and find that there is indeed a phase transition for higher
dimensional $f(R)$ black holes with conformally invariant Maxwell
source.

\end{abstract}

\maketitle
\newpage
\section{Introduction\label{Intro}}

A large volume of observational evidences indicate that our
universe is now experiencing a phase of accelerated expansion.
There are two main approaches for explanation of this cosmic
acceleration. The first approach which try to explain the problem
in the framework of general relativity, requires the existence of
a strange type of energy called ``dark energy" whose gravity is
repulsive and consist an un-clustered component through the
universe. The second approach is to modify general relativity by
adding higher powers of the scalar curvature $R$, the Riemann and
Ricci tensors, or their derivatives in the lagrangian formulation.
Among the latter attempts are Lovelock gravity, braneworld
cosmology, scalar-tensor theories like Brans-Dicke one and also
$f(R)$ theories. A fairly comprehensive review on dark energy
models can be seen in \cite{Cop}. $f(R)$ theories were proved to
be able to mimic the whole cosmological history, from inflation to
the actual accelerated expansion era \cite{Odin,Capo} (see also
\cite{Feli} for a recent review on $f(R)$ theories). Diverse
applications of $f(R)$ theories on gravitation and cosmology have
been also widely studied \cite{Feli}, as well as multiple ways to
observationally and experimentally distinguish them from general
relativity.

When one considers $f(R)$ theory as a modification of general
relativity, it is quite natural to ask about black hole existence
and its features in this theory. It is expected that some
signatures of black holes in $f(R)$ theory may differ from the
expected physical results in Einstein's gravity. Therefore, the
investigation on $f(R)$ black holes is of particular interest.
Some attempts have been done to construct black hole solutions in
$f(R)$ theories (see \cite{Cong,Seb,Ade,Hendi1} and references
therein). In general, in the presence of a matter field, the field
equations of $f(R)$ gravity are complicated and it is not easy to
find exact analytical solutions. However, if one considers the
traceless energy-momentum tensor for the matter field, one can
extract exact analytical solutions from $R+f(R)$ theory coupled to
a matter field \cite{Tae}. Since the energy-momentum tensor of
Maxwell and Yang-Mills fields are traceless only in four
dimensions, therefore black hole solutions from $R+f(R)$ theory
coupled to the matter field were derived only in four dimensions
\cite{Tae}. The studies were also generalized to the four
dimensional charged rotating black holes \cite{Alex}, charged
rotating black string \cite{shey1} and magnetic string solutions
\cite{shey2} in $R+f(R)$-Maxwell theory. However, since the
standard Maxwell energy-momentum tensor is not traceless in higher
dimensions, they failed to derive higher dimensional black
hole/string solutions from $R+f(R)$ gravity coupled to standard
Maxwell field. A natural question then arises: Is there an
extension of Maxwell action in arbitrary dimensions that is
traceless and hence possesses the conformal invariance? The answer
is positive and the conformally invariant Maxwell action was
presented as \cite{Hass1},
\begin{eqnarray}
S_{m}=-\int{d^{n}x\sqrt{-g}(F_{\mu \nu }F^{\mu \nu })^p},
\label{Act1}
\end{eqnarray}
where $p$ is a positive integer, i.e., $p\in \mathbb{N}$. The
associated energy-momentum tensor of the above conformally
invariant Maxwell action is given by
\begin{equation}
T_{\mu \nu}=2 \left(p F_{\mu \eta }F_{\nu }^{\text{ }\eta }
F^{p-1}-\frac{1}{4}g_{\mu \nu } F^{p}\right), \label{T}
\end{equation}
where $F=F_{\alpha \beta}F^{\alpha \beta}$ is the Maxwell
invariant. One can easily check that the above energy momentum
tensor is traceless for $n=4p$. The theory of conformally
invariant Maxwell field is considerably richer than that of the
linear standard Maxwell field and in the special case $(p = 1)$ it
recovers the Maxwell action. It is worthwhile to investigate the
effects of exponent $p$ on the behavior of the solutions and the
laws of black hole mechanics. The motivation is to take advantage
of the conformal symmetry to construct the analogues of the four
dimensional Reissner-Nordstrom (RN) solutions, in higher
dimensions. Recently, the studies on the black object solutions
with a nonlinear Maxwell source in Einstein
\cite{Hass1,Hass2,Hendi2} and Gauss-Bonnet \cite{Mis} gravity have
got a lot of attentions.

In this work we would like to extend the investigation on the
conformally invariant Maxwell field to $f(R)$ gravity. We  will
consider the action (\ref{Act1}) as the matter source of the field
equations in $R+f(R)$ theory with constant curvature scalar. Our
purpose is to find the analogues of the four-dimensional charged
black hole solutions of $R+f(R)$-Maxwell theory \cite{Tae} in
higher dimensional spacetime. In contrast to the higher
dimensional black holes of Einstein gravity with a conformally
invariant Maxwell source presented in \cite{Hass1} which has
vanishing scalar curvature $R=0$, the spacetime we construct here
in $R+f(R)$ gravity coupled to a nonlinear Maxwell field has a
constant scalar curvature $R=R_0$. Our solutions also differ from
higher dimensional RN solutions in that the electric charge term
in the metric coefficient goes as $r^{-(n-2)}$ while in the
standard RN case is $r^{-2(n-3)}$. Also, the electric field in
higher dimensions does not depend on $n$ and goes as electric
field in four dimensions.

This paper is outlined as follows. In Sec. \ref{Field}, we
construct exact spherically symmetric black hole solutions of
$R+f(R)$ theory coupled to a nonlinear Maxwell field in $n=4p$
dimensions and investigate their properties. In Sec. \ref{Therm},
we obtain the conserved and thermodynamic quantities of the
solutions and verify the validity of the first law of black hole
thermodynamics. We also study local stability of the solutions in
this section. We summarize our results in Sec. \ref{Conc}.
%%%%%%%%%%%%%%%%%%%%%%%%%%%%%%%%%%%%%%%%%%%%%%%%%%%%%%%%%%%%%%%%%
\section{Field Equations and solutions\label{Field}}
We consider the action of $R+f(R)$ gravity in $n$-dimensional
spacetime coupled to a conformally invariant Maxwell field
\begin{eqnarray}
S=\int_{\mathcal{M}}d^{n}x\sqrt{-g}\left[ R+f(R)-(F_{\mu \nu
}F^{\mu \nu })^p \ \right], \label{Act}
\end{eqnarray}
where ${R}$ is the scalar curvature, $f(R)$ is an arbitrary
function of scalar curvature, $F_{\mu \nu }=\partial _{\mu }A_{\nu
}-\partial _{\nu }A_{\mu }$ is the electromagnetic field tensor
and $A_{\mu }$ is the electromagnetic potential. The equations of
motion can be obtained by varying action (\ref{Act}) with respect
to the gravitational field $g_{\mu \nu }$ and the gauge field
$A_{\mu }$,
\begin{equation}
{R}_{\mu \nu } \left(1+f^{\prime}(R)\right)-\frac{1}{2}g_{\mu
\nu}(R+f(R))+\left(g_{\mu\nu}\nabla ^{2}-\nabla _{\mu} \nabla
_{\nu}\right)f^{\prime}(R)=T_{\mu \nu} , \label{FE1}
\end{equation}
\begin{equation}
\partial _{\mu}(\sqrt{-g}F^{\mu \nu} F^{p-1}) =0,  \label{FE2}
\end{equation}
where $F=F_{\alpha \beta}F^{\alpha \beta}$ is the Maxwell
invariant and the ``prime'' denotes differentiation with respect
to $R$.  In order to obtain the constant curvature black hole
solution in $f(R)$ gravity  theory coupled to a matter field, the
trace of stress-energy tensor $T_{\mu \nu}$ should be zero
\cite{Tae}. Hence, two candidates for the matter field in four
dimensions are Maxwell and Yang-Mills fields. Since the assumption
of traceless energy-momentum tensor is essential for deriving
exact black hole solutions in $f(R)$ gravity coupled to the matter
field, therefore the solutions exist only for $n=4p$ dimensions.
Assuming the constant scalar curvature $R=R_0=const.$, then the
trace of Eq. (\ref{FE1}) yields
\begin{equation}
{R}_{0} \left(1+f^{\prime}(R_0)\right)-\frac{n}{2}(R_0+f(R_0))=0.
\label{FE3}
\end{equation}
Solving the above equation for $R_0$, gives
\begin{equation}
{R}_{0}=\frac{nf(R_0)}{2f^{\prime}(R_0)+2-n}\equiv\frac{2n}{n-2}{\Lambda_{\rm
f}}<0. \label{R0}
\end{equation}
Substituting the above relation into Eq. (\ref{FE1}), we obtain
the following equation for Ricci tensor
\begin{equation}
{R}_{\mu \nu} (1+f^{\prime}(R_0))-\frac{g_{\mu \nu}}{n}R_0
\left(1+ f^{\prime}(R_0)\right)=T_{\mu \nu}. \label{Ricci}
\end{equation}
We are looking for the $n$-dimensional static spherically
symmetric solutions. Motivated by the metric of higher dimensional
charged black holes in Einstein gravity, we assume the metric has
the following form
\begin{equation}
d{s}^{2}=-N(r )dt^{2}+\frac{dr ^{2}}{N(r )}%
+{r^ 2}d\Omega _{n-2}^{2},  \label{metric}
\end{equation}
where $d\Omega _{n-2}^{2}$ denotes the metric of an unit
$(n-2)$-sphere and $N(r )$ is a functions of $r $ which should be
determined. We are seeking for a purely radial electric solution
which means that the only non-vanishing component of the Maxwell
tensor is $F_{tr}$. In this case, the Maxwell equations
(\ref{FE2}) can be integrated immediately, where, for the
spherically symmetric spacetime (\ref{metric}), all the components
of ${F}_{\mu \nu }$ are zero except ${F}_{tr }$:
\begin{equation}
{F}_{tr }=\frac{q}{r^{\frac{n-2}{2p-1}}},  \label{Ftr0}
\end{equation}
where $q$ is an integration constant. Substituting $n=4p$ in the
above relation, the Maxwell field becomes
\begin{equation}
{F}_{tr }=\frac{q}{r^2}.  \label{Ftr}
\end{equation}
It is important to note that the electric field in higher
dimensions does not depend on $n$ and its value coincides with the
RN solution in four dimensions. Using metric (\ref {metric}) and
the Maxwell field (\ref{Ftr}), one can show that Eq. (\ref{Ricci})
has a solution of the form
\begin{equation}\label{N(r)}
N(r)=1-\frac{2m}{r^{n-3}}+\frac {q^2 }{r^{n-2}}\times
\frac{(-2q^2)^{(n-4)/4}}{\left(1+f^{\prime}(R_0)\right)}-\frac{R_0}{n(n-1)}{r}^{2},
\end{equation}
where $m$ is an integration constant which is related to the mass
of the solution. In four dimension ($n=4$) the solution recovers
the result of \cite{Tae}. In order to have a real solution we
should restrict ourself to the dimensions which are multiples of
four, i.e., $n=4,8, 12,....$, which means that $p$ should be only
positive integer, as we mentioned already.

\begin{figure}[tbp]
\epsfxsize=7cm  \centerline{\epsffile{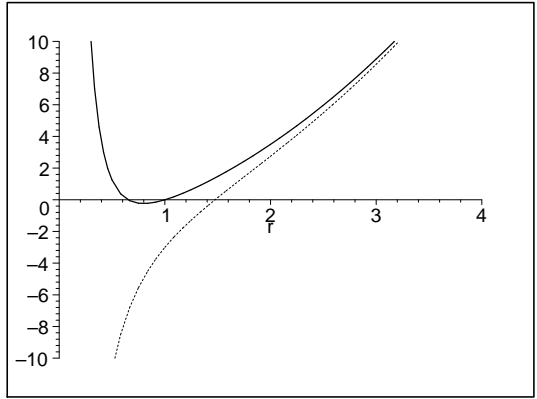}} \caption{The
function $N(r)$ versus $r$ for $m=2$, $n=4$, $q=1$  and $R_0=-12$.
$f'(R_0)=-0.5$ (bold line) and $f'(R_0)=-2$ (dashed line).}
\label{figure1}
\end{figure}

\begin{figure}[tbp]
\epsfxsize=7cm \centerline{\epsffile{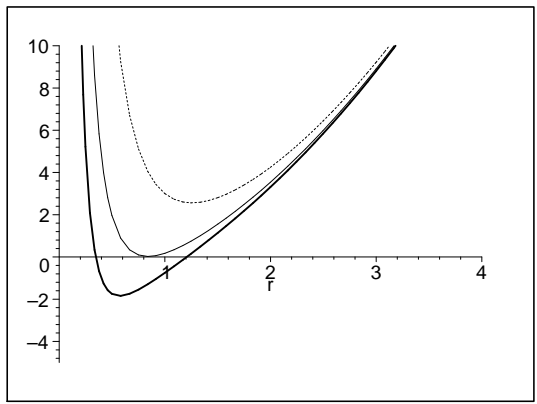}} \caption{The
function $N(r)$ versus $r$ for $m=2$, $q=1$,  $n=4$ and $R_0=-12$.
$f'(R_0)=-0.2$ (bold line), $f'(R_0)=-0.54$ (continuous line) and
$f'(R_0)=-0.8$ (dashed line).} \label{figure2}
\end{figure}

Next we study the physical properties of the solutions. To do
this, we first look for the curvature singularities. A simple
calculation shows that the Kretschmann scalar $R_{\mu \nu \lambda
\kappa }R^{\mu \nu \lambda \kappa }$ diverges at $r=0$, it is
finite for $r\neq 0$ and is proportional to $R_0^2$ as
$r\rightarrow \infty $. Therefore, there is a curvature
singularity located at $r=0$. As one can see from Eq.
(\ref{N(r)}), the solution is ill-defined for
$f^{\prime}(R_0)=-1$. Let us consider the cases with
$f^{\prime}(R_0)>-1$ and $f^{\prime}(R_0)<-1$ separately. In the
first case where $f^{\prime}(R_0) >-1$,  we have a black hole
solution (see Fig. \ref{figure1}). Indeed, for
$1+f^{\prime}(R_0)>0$ and $R_0<0$ black hole can have two inner
and outer horizons, an extreme black hole or a naked singularity
provided the parameters of the solutions are chosen suitably (see
Fig. \ref{figure2}). In the latter case where
$f^{\prime}(R_0)<-1$, we encounter a cosmological horizon for
$R_0<0$. Indeed, in this case the signature of the spacetime
changes and the conserved quantities such as mass become negative,
as we will see in the next section, thus this is not a physical
case and we rule it out from our consideration.

It is apparent that the spacetime described by solution
(\ref{N(r)}) is asymptotically AdS provided we define
$R_0=-n(n-1)/l^2$. However, the solution presented here differ
from the standard higher dimensional Reissner-Nordstrom AdS
(RNAdS) solutions since the electric charge term in the metric
coefficient goes as $r^{-(n-2)}$ while in the standard RNAdS case
is $r^{-2(n-3)}$. In four dimensions, with the following
replacement
\begin{eqnarray}\label{rep}
&&\frac{\alpha q^2}{\left(1+f^{\prime}(R_0)\right)} \rightarrow Q^2\\
&&{R_0}\rightarrow 4\Lambda
\end{eqnarray}
the solution reduces to standard RNAdS black holes for
$\Lambda=-3/l^2$.

We use the subtraction method of Brown and York (BY) \cite{BY} to
calculate the quasilocal mass of the charged $f(R)$ black hole.
Such a procedure causes the resulting physical quantities to
depend on the choice of reference background. In order to use the
BY method one should write the metric in the following form
\begin{equation}
ds^{2}=-W(r)dt^{2}+\frac{d{r}^{2}}{V({r})}+{r}%
^{2}d\Omega_{n-2} ^{2}. \label{Mets}
\end{equation}
Since metric (\ref{metric}) has the above form, it is sufficient
to choose the background metric to be the metric (\ref{Mets}) with
\begin{equation}
W_{0}({r})=V_{0}({r})=N_{0}(r)=1+\frac{r^2}{l^2}
\end{equation}
Where we have defined $R_0=-n(n-1)/l^2$ which show that the
solutions are asymptotically AdS as we mentioned. It is well-known
that the Ricci scalar for AdS spacetime should have this value
(see e.g. \cite{Weinberg}). To compute the conserved mass of the
spacetime, we choose a timelike Killing vector field $\xi $ on the
boundary surface ${\cal B}$ of the spacetime (\ref {Mets}). Then
the quasilocal conserved mass can be written as
\begin{equation}
{\cal M}=\frac{1}{8\pi }\int_{{\cal B}}d^{n-2}x \sqrt{\sigma
}\left\{ \left( K_{ab}-Kh_{ab}\right) -\left(
K_{ab}^{0}-K^{0}h_{ab}^{0}\right) \right\} n^{a}\xi ^{b},
\end{equation}
where $\sigma $ is the determinant of the metric of the boundary
${\cal B}$, $K_{ab}^{0}$ is the extrinsic curvature of the
background metric  and $n^{a}$ is the timelike unit normal vector
to the boundary ${\cal B}$. In the context of counterterm method,
the limit in which the boundary ${\cal B}$ becomes infinite
(${\cal B}_{\infty }$) is taken, and the counterterm prescription
ensures that the action and conserved charges are finite. Thus, we
obtain the mass through the use of the above subtraction method of
BY as
\begin{equation}
{M}=\frac{(n-2) \Omega _{n-2}}{8\pi }m\left[
1+f^{\prime}(R_0)\right], \label{M}
\end{equation}
where $\Omega _{n-2}$ is the volume of the unit $(n-2)$-sphere. In
the limiting case  ($f^{\prime}(R_0)=0$), this expression for the
mass reduces to the mass of the $n$-dimensional AdS black hole.
%%%%%%%%%%%%%%%%%%%%%%%%%%%%%%%%%%%%%%%%%%%%%%%%%%%%%%%%
\section{Thermodynamics of Charged $f(R)$ Black holes \label{Therm}}
In this section we are going to explore thermodynamics of higher
dimensional charged $f(R)$ black holes. The Hawking temperature of
the black holes can be easily obtained by requiring the absence of
conical singularity at the horizon in the Euclidean sector of the
black hole solutions. One obtains the associated temperature with
the outer event horizon $r=r_{+}$ as
\begin{eqnarray}\label{T}
T&=&\frac{1}{4\pi}\left(\frac{d N(r)}{dr}\right)_{r=r_{+}}
=\frac{[1+f^{\prime}(R_0)] \left(2
r^2(n-1)+2l^2(n-3)\right)+(-2q^2)^{n/4} r^{2-n} l^2}{\pi l^2
r[1+f^{\prime}(R_0)] }.
\end{eqnarray}
where we have used equation $N(r_{+})=0$ for omitting the mass
parameter $m$ from temperature expression. Next, we calculate the
entropy of the black hole. Let us first give a brief discussion
regarding the entropy of the black hole in $f(R)$ gravity. To this
aim, we follow the arguments presented in \cite{Brevik}. If one
use the Noether charge method for evaluating the entropy
associated with black hole solutions in $f(R)$ theory with
constant curvature, one finds \cite{Cong}
\begin{equation}
{S}=\frac{A}{4G}f^{\prime}(R_0), \label{S0}
\end{equation}
where $A=4 \pi r_{+}^2$ is the horizon area. As a result, in
$f(R)$ gravity, the entropy does not obey the area law and one
obtains a modification of the area law. Inspired by the above
argument, for the $n$-dimensional charged black hole solutions in
$R+f(R)$ gravity, we find the entropy as
\begin{equation}
{S}=\frac{r_+^{n-2}\Omega _{n-2}}{4} \left[1+f'(R_0)\right].
\label{S}
\end{equation}
The charge of  conformally invariant $f(R)$ black holes can be
found by calculating the flux of the electric field at infinity,
yielding
\begin{equation}
{Q}=\frac{ n (-2)^{(n-4)/4}q^{(n-2)/2} \Omega _{n-2}}{16\pi
\sqrt{1+f'(R_0)}}. \label{Q}
\end{equation}
The electric potential $\Phi$, measured at infinity with respect
to the horizon, is defined by
\begin{equation}
\Phi=A_{\mu }\chi ^{\mu }\left| _{r\rightarrow \infty }-A_{\mu
}\chi ^{\mu }\right| _{r=r_{+}},  \label{Pot}
\end{equation}
where $\chi=\partial_{t}$ is the null generator of the horizon. We
find
\begin{equation}
\Phi=\frac{q}{ r_{+}}\sqrt{1+f'(R_0)}. \label{Pot}
\end{equation}
Then, we investigate the validity of the first law of
thermodynamics for higher dimensional charged $f(R)$ black hole.
For this purpose, we obtain a Smarr-type formula, namely the mass
$M$ as a function of extensive quantities $S$, and $Q$. Using the
expression for the mass, the entropy and the charge given in Eqs.
(\ref{M}), (\ref{S}) and (\ref{Q}) and the fact that $N(r_{+})=0$,
we find
\begin{eqnarray}
M(S,Q)&=&\frac{n-2}{16 \pi
l^2}\left(\frac{4S}{1+f'(R_0)}\right)^{-1/(n-2)}\left[
[1+f'(R_0)]\left(\frac{4S}{1+f'(R_0)}\right)\times \right.
\nonumber
\\
&&
\left.\left(l^2+\left(\frac{4S}{1+f'(R_0)}\right)^{2/(n-2)}\right)
+l^2q^2(-2q^2)^{(n-4)/4}\right]. \label{Msmar}
\end{eqnarray}
where $q$ is a function of $Q$ according to Eq. (\ref{Q}). One may
then regard the parameters $S$, and $Q$ as a complete set of
extensive parameters for the mass $M(S,Q)$ and define the
intensive parameters conjugate to $S$ and $Q$. These quantities
are the temperature and the electric potential
\begin{equation}
T=\left( \frac{\partial M}{\partial S}\right) _{Q},\ \ \Phi=\left( \frac{\partial M%
}{\partial Q}\right) _{S}.  \label{Dsmar}
\end{equation}
Numerical calculations show that the intensive quantities
calculated by Eq. (\ref{Dsmar}) coincide with Eqs. (\ref{T}) and
(\ref{Pot}). Thus, the thermodynamics quantities we obtained in
this section satisfy the first law of black hole thermodynamics
\begin{equation}
dM = TdS+\Phi d{Q}.
\end{equation}
In this way we constructed all conserved and thermodynamic
quantities of charged $f(R)$ black holes and verified the validity
of the first law of thermodynamics on the event horizon.
\begin{figure}[bp]
\epsfxsize=7cm \centerline{\epsffile{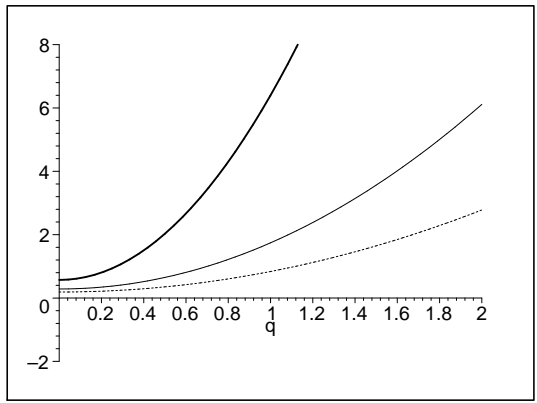}} \caption{The
function $(\partial ^2M/\partial S^2)_{Q}$ versus $q$ for $l=1$,
$n=4$, $r_{+}=0.8$. $f'(R_0)=-0.5$ (bold line), $f'(R_0)=0$
(continuous line) and $f'(R_0)=0.5$ (dashed line).}
\label{figure3}
\end{figure}

\begin{figure}
\epsfxsize=7cm \centerline{\epsffile{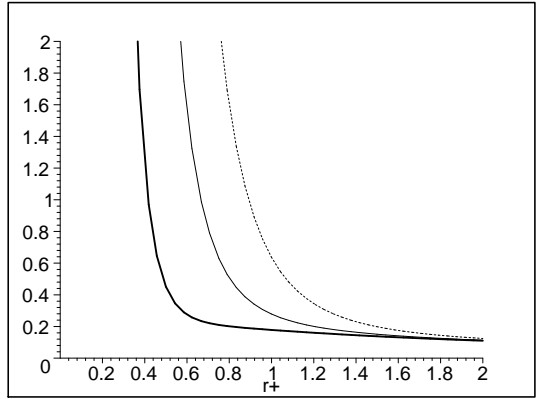}} \caption{The
function $(\partial ^2M/\partial S^2)_{Q}$ versus $r_{+}$ for
$l=1$, $f'(R_0)=1$ and $n=4$. $q=0.4$ (bold line), $q=1$
(continuous line) and $q=2$ (dashed line).} \label{figure4}
\end{figure}
\begin{figure}
\epsfxsize=7cm \centerline{\epsffile{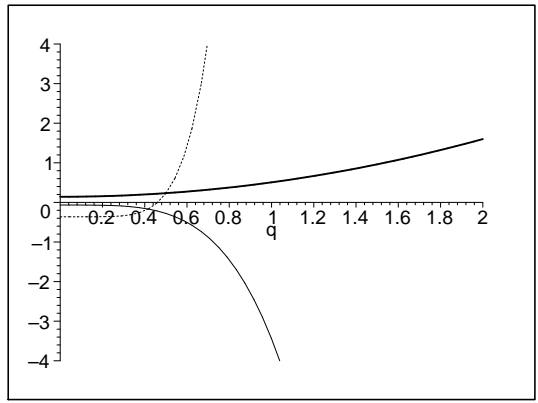}} \caption{The
function $(\partial ^2M/\partial S^2)_{Q}$ versus $q$ for $l=1$,
$r_{+}=0.8$ and $f'(R_0)=1$. $n=4$ (bold line), $n=8$ (continuous
line) and $n=12$ (dashed line).} \label{figure5}
\end{figure}
\begin{figure}
\epsfxsize=7cm \centerline{\epsffile{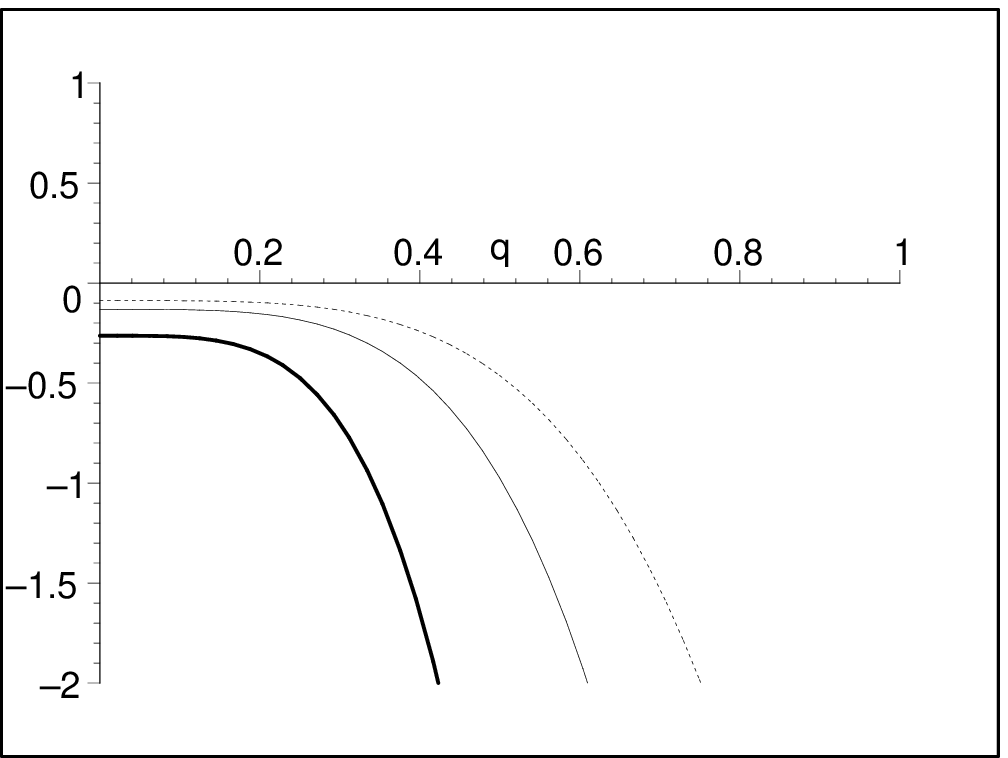}} \caption{The
function $(\partial ^2M/\partial S^2)_{Q}$ versus $q$ for $l=1$,
$n=8$ and $r_{+}=0.8$. $f'(R_0)=-0.5$ (bold line), $f'(R_0)=0$
(continuous line) and $f'(R_0)=0.5$ (dashed line).}
\label{figure6}
\end{figure}
\begin{figure}
\epsfxsize=7cm \centerline{\epsffile{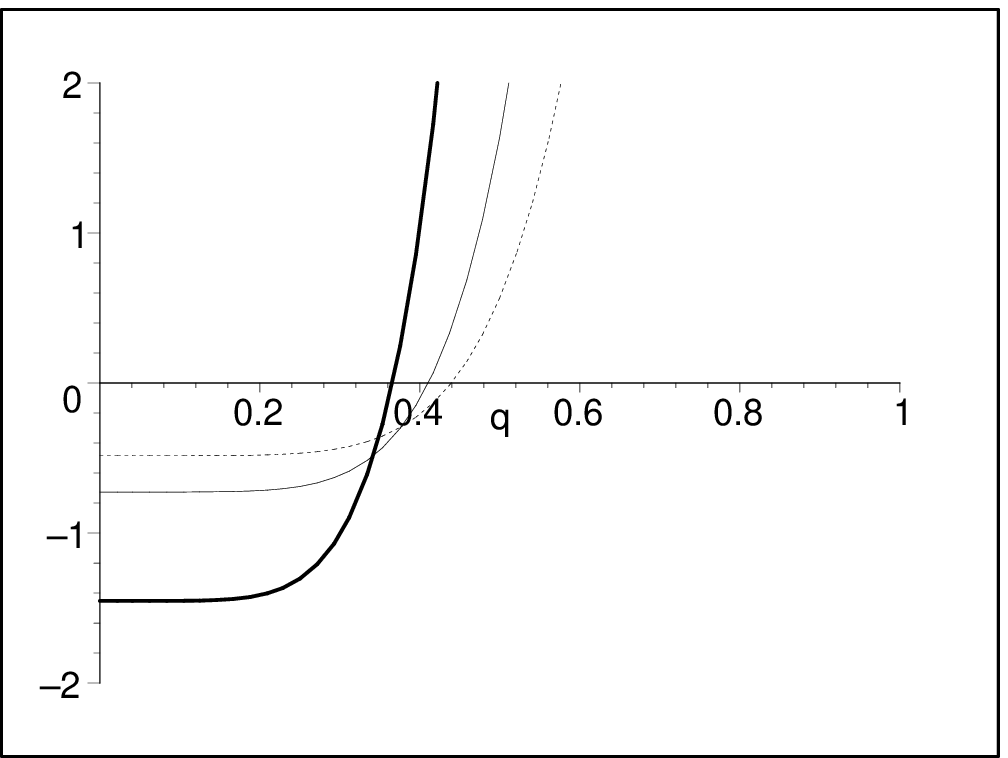}} \caption{The
function $(\partial ^2M/\partial S^2)_{Q}$ versus $q$ for $l=1$,
$n=12$ and $r_{+}=0.8$. $f'(R_0)=-0.5$ (bold line), $f'(R_0)=0$
(continuous line) and $f'(R_0)=0.5$ (dashed line).}
\label{figure7}
\end{figure}
Finally, we study the local stability of the charged $R+f(R)$
black holes in the presence of conformally invariant Maxwell
source. In the canonical ensemble, the positivity of the heat
capacity $C_{Q} = T/(\partial^2 M/\partial S^2)_{Q}$ and therefore
the positivity of $(\partial^2 M/\partial S^2)_{Q}$ is sufficient
to ensure the local stability. We have shown the behavior of
$(\partial ^{2}M/\partial S^{2})_{Q}$\ as a function $q $ and
$r_{+}$ for different value of $n$, $q$ and $f'(R_0)$ in figures
3-7 where we have taken $R_0=-n(n-1)/l^2$. From figures 3 and 4 we
see that the system is always thermally stable in four dimensions
for different value of the parameters. However, in higher
dimensions the system has an unstable phase as one can see from
Fig. 5. As an example, we find that in $8$-dimensions black holes
are always unstable and there is no phase transition (see Fig. 6),
while in $12$-dimensions the system has a transition from unstable
phase to stable phase (see Fig. 7).
\section{Summary and Discussion\label{Conc}}
In general the field equations of $f(R)$ gravity coupled to a
matter field are complicated and it is not easy to construct exact
analytical solutions. Recently, it was shown \cite{Tae} that by
assuming a traceless energy-momentum tensor for the matter field
as well as the constant curvature scalar, one can extract some
analytical black hole solutions in $R+f(R)$ theory coupled to a
matter field. Two examples for the traceless $T_{\mu \nu}$ in four
dimensions are Maxwell and Yang-Mills fields \cite{Tae}. However,
the energy-momentum tensor of Maxwell field is not traceless in
higher dimensions. Seeking for a traceless energy-momentum tensor
in arbitrary dimensions, the authors of \cite{Hass1} found a
conformally invariant nonlinear Maxwell action which its
energy-momentum tensor is traceless in $n=4p$ dimensions where $p$
is a positive integer. They also studied the black hole solutions
in Einstein gravity with nonlinear Maxwell field \cite{Hass1}.

In this paper we obtained a class of higher dimensional black
holes from $R+f(R)$ gravity with conformally invariant Maxwell
source. The two key assumptions in finding these solutions are:
(i) the constant scalar curvature $R=R_{0}$ and (ii) the traceless
energy momentum tensor. These solutions are similar to higher
dimensional RNAdS black holes with appropriate replacement of the
parameters, but only exist in dimensions which are multiples of
four. Besides, the solutions presented here differ from higher
dimensional RNAdS black holes in two features. First, the electric
charge term in the metric coefficient goes as $r^{-(n-2)}$ while
in the standard RNAdS case is $r^{-2(n-3)}$. Second, the electric
field in higher dimensions does not depend on $n$ and goes as
electric field in four dimensional RNAdS black holes. Our
solutions also differ from the higher dimensional black holes of
Einstein gravity with a conformally invariant Maxwell source
\cite{Hass1} in that they have vanishing scalar curvature $R=0$,
while the obtained solutions here in $R+f(R)$ gravity coupled to a
nonlinear Maxwell field have a constant  curvature scalar $R=R_0$.
In addition, the conserved and thermodynamic quantities computed
here depend on function $f'(R_{0})$ and differ completely from
those of Einstein theory in AdS spaces. Clearly the presence of
the general function $f'(R_{0})$ changes the physical values of
conserved and thermodynamic quantities. Furthermore, unlike
Einstein gravity, for the black hole solutions obtained here in
$f(R)$ gravity, the entropy does not obey the area law.

After studying the physical properties of the solutions, we
computed the mass, charge, electric potential and temperature of
the black holes. We also found the entropy expression which does
not obey the area law for the $f(R)$ black holes.  We obtained a
Smarr-type formula for the mass, and verified that the conserved
and thermodynamics quantities satisfy the first law of black hole
thermodynamics. We also studied the phase behavior of the higher
dimensional charged $f(R)$ black holes. We found that the system
is always thermally stable in four dimensions, while in higher
dimensions, there is a phase transition in the presence of the
conformally invariant Maxwell field in $R+f(R)$ theory with
constant curvature.

Finally, we would like to mention that the $n$-dimensional charged
$f(R)$ black hole solutions obtained here are static. Thus, it
would be interesting if one can construct charged rotating black
brane/hole solutions from $R+f(R)$ theory in the presence of
conformally invariant Maxwell source. These issues are now under
investigation and will be appeared elsewhere.
%%%%%%%%%%%%%%%%%%%%%%%%%%%%%%%%%%%%%%%%%%%%%%%%%%%%%%%%%%%%%%%%%%%
\acknowledgments{This work has been supported financially by
Center for Excellence in Astronomy and Astrophysics (CEAA-RIAAM),
Maragha, under research project number 1/2718.}
%%%%%%%%%%%%%%%%%%%%%%%%%%%%%%%%%%%%%%%%%%%%%%%%%%%%%%%%%%%%%%%%%%%

\end{document}